\documentclass{aastex631}
\usepackage{bm}
\usepackage{ amssymb }
\usepackage{cancel}

\begin{document}

\title{Including Neutrino-driven Convection into the Force Explosion Condition to Predict Explodability of Multi-dimensional Core-collapse Supernovae (FEC+)}

\correspondingauthor{Mariam Gogilashvili}
\email{mgogilashvili@fsu.edu}

\author[0000-0002-6944-8052]{Mariam Gogilashvili}
\affiliation{Florida State University, Tallahassee, Florida, 32306, USA}
\affiliation{CCS-2, Computational Physics and Methods, Los Alamos National Laboratory, Los Alamos NM 87544, USA}

\author[0000-0003-1599-5656]{Jeremiah W. Murphy}
\affiliation{Florida State University, Tallahassee, Florida, 32306, USA}

\author[0000-0001-6432-7860]{Jonah M. Miller}
\affiliation{CCS-2, Computational Physics and Methods, Los Alamos National Laboratory, Los Alamos NM 87544, USA}
\affiliation{Center for Theoretical Astrophysics, Los Alamos National Laboratory, Los Alamos, NM, USA}
\affiliation{Center for Nonlinear Studies, Los Alamos National Laboratory, Los Alamos, NM, USA}

\begin{abstract}
Most massive stars end their lives with core collapse. However, it is not clear which explode as a Core-collapse Supernova (CCSN), leaving behind a neutron star and which collapse to black hole, aborting the explosion. One path to predict explodability without expensive multi-dimensional simulations is to develop analytic explosion conditions.  These analytic explosion conditions also provide a deeper understanding of the explosion mechanism and they provide some insight as to why some simulations explode and some do not.  The analytic force explosion condition (FEC) reproduces the explosion conditions of spherically symmetric CCSN simulations.  In this followup manuscript, we include the dominant multi-dimensional effect that aids explosion, neutrino driven convection, into the FEC.  This generalized critical condition (FEC+) is suitable for multi-dimensional simulations and has potential to accurately predict explosion conditions of two- and three-dimensional CCSN simulations.  We show that adding neutrino-driven convection reduces the critical condition by $\sim 30\%$, which is consistent with previous multi-dimensional simulations. \end{abstract}

\section{Introduction} \label{sec:intro}
At the end of their lives, massive stars undergo core collapse. Some explode as core-collapse supernovae (CCSNe), leaving behind a neutron star (NS) \citep{LI2011, HORIUCHI2011}, some of them fail to explode and collapse to a black hole (BH) \citep{FISCHER2009, OCONNOR2011}, and a small subset may explode and produce a BH through fallback \citep{ERTL2016,SUKHBOLD2016,Ebinger2019}. Understanding how and which massive stars explode has been a major challenge for decades. Predicting which massive stars explode is key to fully understanding  nucleosynthesis \citep{WOOSLEY2002, DIEHL2021}, neutron star and black hole formation \citep{BAADE1934A, BURROWS1986, FISCHER2009, OCONNOR2011}, and neutron star and black hole distributions.

The CCSN mechanism is nonlinear and involves hydrodynamics, general relativity, radiation transport, and nuclear physics. Given this complexity in physics, one approach to understand the CCSN mechanism is to run expensive multi-dimensional simulations \citep{MULLER2017,RADICE2017,VARTANYAN2021}. 

Another approach is to develop analytic explosion conditions; these can provide a deeper understanding of the explosion mechanism and quantify how close simulations are to explosion.  Over the years, many have tried to develop a theory for successful explosions \citep{THOMPSON2000, PEJTA2012, RIVES2018, RAIVES2021,SUMMA2016, SUMMA2018, YAMASAKI2005, JANKA2012, JANKA2016, KESHET2012,OCONNOR2011, WANG2022, Boccioli2023}. One of the first was the semi-analytic critical curve of \citet{BURROWS1993}. \citet{BURROWS1993} assumed that the CCSN problem can be considered as a steady-state, boundary value problem and solved the ODEs with the inner boundary being at the NS and the outer boundary at the shock. They considered a two-dimensional parameter space of the neutrino luminosity $L_\nu$ and the mass accretion rate, $\dot{M}$ and found that there is a critical curve in this parameter space below which there are stalled-shock solutions, but above this curve they found no stalled-shock solutions. They suggested that the region above this curve is the explosive region. Later, \citet{MURPHY2008} showed that the semi-analytic critical curve of \citet{BURROWS1993} is consistent with 1D simulations. To explain the physical origin of the critical curve of \citet{BURROWS1993}, \citet{MURPHY2017} found that explosions occur when the integral of the momentum equation, $\Psi$ is greater than $0$.  From this observation, \citet{MURPHY2017} suggested a path toward deriving an analytic explosion condition.

Inspired by this suggestion, \citet{Gogilashvili2022} analytically derived an analytic critical condition for spherically symmetric explosions, called it the Force Explosion Condition (FEC) due to the fact that it is indeed the imbalance of forces that determine the explodability. Strikingly, they found that the explosion condition only depends upon two dimensionless parameters. \citet{Gogilashvili2022} first tested the consistency of the FEC with the semi-analytic solutions of \citet{BURROWS1993}. Then they checked the validity of the FEC using the light-bulb simulations \citep{Gogilashvili2022} as well as using 1D radiation hydrodynamics simulations with the appropriate neutrino transport \citep{Gogilashvili2023}. In both cases, they find that the FEC accurately predicts the explodability of 1D simulations and is a good diagnostic of how far away a simulation is from explosion. While these successes of the FEC are encouraging, this simple model assumes spherical symmetry and is valid only for spherical explosions.  Yet, a significant body of work shows that multi-dimensional effects such as neutrino-driven convection aids neutrinos in driving the explosion 
\citep{Bethe1990,Janka1996,Foglizzo2007,MABANTA2018}.

In this followup manuscript, we generalize the FEC by including multi-dimensional effects in the model.  In particular, we incorporate neutrino-driven convection and turbulent dissipation into the derivation of \citet{Gogilashvili2022}. We show that adding neutrino-driven convection reduces the critical condition around $\sim 30\%$ which is consistent with previous results from multi-dimensional simulations \citep{MURPHY2008, MURPHY2011, MABANTA2018}. 

The structure of the manuscript is as follows: In section~\ref{ssec:FEC}, we review the spherical model of the FEC of \citet{Gogilashvili2022}. In section~\ref{ssec:FEC+derivation}, we derive the generalized FEC (FEC+). In section~\ref{ssec:Applications}, we review the applications of the FEC+ for different convection models. In particular, section~\ref{ssec:MFC} derives applications of the FEC+ for the integral convection model of \citet{MABANTA2019} and section~\ref{ssec:stir}- for the mixing-length-theory model of Supernova Turbulence In Reduced-dimensionality (STIR) \citep{Couch2020}. Lastly, in section~\ref{sec:discussion}, we discuss and summarize the main conclusion and provide next steps for using the FEC+.

\section{The Force Explosion Condition with Neutrino-Driven Convection} \label{sec:FEC+}
\subsection{A Brief Review of the Force Explosion Condition (FEC)}\label{ssec:FEC}

\citet{Gogilashvili2022} derived an analytic model for the explosion condition. The derivation is complete and self-consistent in that it begins with the fundamental equations of hydrodynamics and is explicit about all approximations.  The force explosion condition (FEC) considers the balance of integrated forces and associated boundary conditions \citep{MURPHY2017}.  Using dimensional analysis, \citet{Gogilashvili2022} also demonstrate that the explosion condition only depends upon two dimensionless parameters...as opposed to four dimension-full parameters.  

The force explosion condition is $\tilde{L}_\nu\tau_g-a\tilde{\kappa}\geq b$.  $\tilde{L}_{\nu} \tau_g$ is the net neutrino power deposited in the gain region normalized by the accretion power:  $\tilde{L}_\nu\tau_g=L_{\nu} \tau_g R_{\rm NS}/ ( G \dot{M} M_{\rm NS})$. $\tilde{\kappa}$ is the dimensionless neutrino opacity that parameterizes the neutrino optical depth in the accreted matter near the neutron star surface: $\tilde{\kappa}=\kappa \dot{M} / \sqrt{G M_{\rm NS} R_{\rm NS}}$. The coefficient $a$ is the difference of the post-shock pressure and the ram pressure of $\dot{M}$, and $b$ is related to gravity. \citet{Gogilashvili2022} analytically estimated these coefficients to be $a\sim 0.05$ and $b\sim 0.52$. 

\citet{Gogilashvili2022} tested the accuracy of the FEC in two ways. The first approach showed that the FEC is a more general expression of the semi-analytic condition of \citet{BURROWS1993}.  Furthermore, this comparison led to a numerical fit for $a = 0.06$ and $b=0.38$, which are remarkably close to the analytic estimates. The second test compares, the FEC with  one-dimensional (1D) simulations that use local neutrino heating and cooling approximations (1D light-bulb simulations).  The FEC accurately predicts the explodability of 1D Light-Bulb simulations (Figure 10 of \citet{Gogilashvili2022}). 

In a followup manuscript, \citet{Gogilashvili2023} compared the FEC with 1D simulations that use actual neutrino transport.  To make this comparison, they also proposed small modifications to the FEC. For example, replacing $L_\nu \tau_g$ with the net neutrino heating rate in the gain region, $\dot{Q}_\nu$, is a more accurate measure of neutrino power for neutrino transport. They tested the FEC using \textit{GR1D}, which simulates the spherically symmetric CCSN using GR and neutrino transport\citep{OCONNOR2010,OCONNOR2015}. The FEC accurately predicts the explosion condition for these spherically symmetric CCSN simulations that use neutrino transport (Figure 5 of \citet{Gogilashvili2023}). 

To further generalize the FEC for two- and three-dimensional simulations, \citet{Gogilashvili2022} and \citet{Gogilashvili2023} proposed a way to add multi-dimensional effects such as neutrino-driven convection and turbulent dissipation to the analytic model of the FEC.  Section~\ref{ssec:FEC+derivation} presents the derivation of FEC+, which is the FEC with neutrino-driven convection.  

\subsection{The FEC+}\label{ssec:FEC+derivation}
To develop a FEC for multi-dimensional simulations, one needs to include multi-dimensional effects in the simple spherical model. Since neutrino-driven convection is the dominant multi-dimensional instability that aids explosion \citep{MABANTA2018}, the following derivation incorporates neutrino-driven convection.  Before adding a specific convection model, this section presents a more general derivation of FEC+ that accommodates any convective or turbulent, mean-field model. Much of the following derivation follows the procedure of  \citet{MABANTA2019} (section~\ref{ssec:MFC}). The derivation starts with the fundamental equations of hydrodynamics; using Einstein notation, they are:

\begin{equation}\label{eq:cont}
    \rho_{,t}+(\rho u^i)_{;i}=0 \, ,
\end{equation} 
\begin{equation}\label{eq:mom}
   (\rho u_i)_{,t}+(\rho u_i u^j+\delta^j_i P)_{;j}=-\rho \Phi_{,i} \, ,
\end{equation} 
\begin{equation}\label{eq:energy}
    (\rho E)_{,t}+(\rho u^j[h+\frac{u^2}{2}])_{;j}=-\rho u^j \Phi_{,j}+\rho q \, ,
\end{equation} 
where $\rho$ is the mass density, $u$ is the fluid velocity, $P$ is the pressure, $h$ is the enthalpy, $\Phi$ is the gravitational potential, and $q$ is net heating, which includes local specific neutrino heating, $q_\nu=\frac{L_\nu \kappa}{4\pi r^2}$, and cooling, $q_c=C_0\big(\frac{T}{T_0}\big)^6$ \citep{JANKA2001}.  

Reynolds decomposition separates the fluid variables into the background (subscript $0$) and convective parts (superscript '), $\chi=\chi_0+\chi'$, where $\chi$ is a fluid variable. Following the same technique and assumptions, the Reynolds decomposed equations are (see details in appendix \ref{Appendix}  and in \citet{MABANTA2019}):

\begin{equation}\label{eq:Rc}
    \rho_{0_,t}+(\rho_0 u_0^i)_{;i}=0 \, , 
\end{equation}  
\begin{equation}\label{eq:Rm}
    (\rho_0 u_{0_i})_{,t}+\left \langle \rho'u'_i\right \rangle _{,t} + (\rho_0 u_{0i} u_0^j+\delta^j_i P_0)_{;j}+\left \langle \rho_0 R^j_i\right \rangle _{;j}=-\rho_0 \Phi_{,i} \, ,
\end{equation}
\begin{equation}\label{eq:Re}
    \left \langle \rho E\right \rangle _{,t}+(\rho_0 u_0^j(\varepsilon_0+\frac{u_0^2}{2}))_{;j}+(u_0^j P_0)_{;j}=-(\rho_0 u_0^j \Phi_{,j})+\rho_0 q+W_b-\left \langle F^j_I\right \rangle _{;j} \, ,  
\end{equation}
where $R^i_j=u'^iu'_j$ is the Reynolds stress, $F_I^i=\rho\varepsilon u'^i$  is the perturbed internal energy flux, and $W_b=-\left \langle \rho'u'^j\right \rangle \Phi_{,j}$ is the work done by buoyant driving.
To find an analytic explosion condition given these equations, we follow the prescriptions outlined in \citet{MURPHY2017} and  \citet{Gogilashvili2022}.  The first step is to assume steady-state; this transforms the evolution equations into a boundary value problem.  The lower boundary is the NS ``surface'' or electron neutrinosphere, and the upper boundary is the shock.  The second assumption is to assume spherical symmetry for both the background quantities and the convective flow.  To facilitate analytic and algebraic solutions, we integrate the steady-state equations from the NS surface to the shock.

Boundary conditions and an equation of state (EoS) complete these equations.  To facilitate analytic solutions, we assume that the pre-shock material is in free fall and pressureless ($P_{0+}=0$). We also assume that $\dot{M}$ is constant. Moreover, we assume that the Reynolds stress for pre-shock material is negligible ($\left \langle R^r_r\right \rangle \approx 0$). The resulting shock jump conditions are:
\begin{equation}\label{eq:jump1}
    \rho_{0+} u_{0+}=\rho_{0-}u_{0-}
\end{equation}
\begin{equation}\label{eq:jump2}
    \rho_{0+}u_{0+}^2=\rho_{0-}u_{0-}^2+P_{0-}+\rho_{0-}\left \langle R^r_r\right \rangle _-
\end{equation}
\begin{equation}\label{eq:jump3}
    \dot{M}(\gamma\varepsilon+\frac{u^2}{2}+\Phi)_-+4\pi R_s^2 F_{I_{-\epsilon}}=\dot{M}(\gamma\varepsilon+\frac{u^2}{2}+\Phi)_+
\end{equation}
A  $\gamma-$law EoS, $P=(\gamma-1)\rho\varepsilon$ further facilitates analytic solutions.  Eqs.~(\ref{eq:Rc})-(\ref{eq:Re}) along with the EoS and boundary conditions (eqs.~(\ref{eq:jump1})-(\ref{eq:jump3})) fully describe the evolution of spherically symmetric background flow with neutrino-driven convection. 

\citet{Gogilashvili2022} derived the FEC condition from the (im)balance of integral forces in the momentum equation.  Here, we derive the integral forces including Reynolds-decomposed terms.  The resulting balance of forces, $\Psi$, is
\begin{eqnarray}\label{eq:psi}
    \Psi=&-&\int^{R_s-\epsilon}_{R_{\rm NS}}{\frac{1}{r^2}\frac{\partial}{\partial r}(r^2 \rho_0 u_{0r}^2) dV} - \int^{R_s-\epsilon}_{R_{\rm NS}}{\frac{\partial P_0}{\partial r}dV}-\int^{R_s-\epsilon}_{R_{\rm NS}}{\rho_0\frac{\partial\Phi}{\partial r}dV}\\ \nonumber
&-&\int^{R_s-\epsilon}_{R_{\rm NS}}[\frac{1}{r^2}\frac{\partial}{\partial r}\left \langle r^2\rho_0 R^r_r\right \rangle +\frac{1}{r}\left \langle \rho_0R^r_r\right \rangle ]dV 
\end{eqnarray}
The first three terms on the right side of eq.~(\ref{eq:psi}) are the same terms as in \citet{Gogilashvili2022}. Integrating the last two terms, on the other hand, requires additional assumptions for the density profile and Reynolds stress. We scale the density to the density at the surface of the NS, $\rho=\rho_{\rm NS} f(x)$, where $x = r/R_{\rm NS}$. Previous simulations show that the average radial component of the Reynolds stress is zero at the surface of the NS, $\left \langle R^r_r\right \rangle _{\rm NS}\approx0$ (see \citet{MURPHY2013}) and for simplicity, it is reasonable to assume that the Reynolds stress is constant in the gain region, $\left \langle R^r_r\right \rangle =\left \langle R^r_r\right \rangle _-$. With these assumptions, the last two terms are:
\begin{equation}
    \int^{R_s-\epsilon}_{R_{\rm NS}}{\frac{1}{r^2}\frac{\partial}{\partial r}(r^2 \rho_0\left \langle R^r_r\right \rangle ) dV}=4\pi R_s^2 \rho_{0-}\left \langle R^r_r\right \rangle _- \, ,
\end{equation}
and
\begin{equation}
    \int^{R_s-\epsilon}_{R_{\rm NS}}\frac{1}{r}\left \langle \rho_0R^r_r\right \rangle dV=4\pi\rho_{\rm NS}\left \langle R_r^r\right \rangle _- R_{\rm NS}^2\int^{x_s-\epsilon}_{x_{g}}{f(x)x }dx \, ,
\end{equation}
where $x_s$ and $x_g$ are dimensionless shock radius and gain radius respectively.

Including the jump conditions (eqs.~(\ref{eq:jump1})\&(\ref{eq:jump2})), denoting the shock compression ratio as $\beta=\rho_-/\rho_+$, and dropping subscripts "0" and "r", $\Psi$ becomes:
\begin{eqnarray}
    \Psi=-R_s^2[\rho_- u_-^2+\rho_-\left \langle R^r_r\right \rangle _-]-\left \langle r^2\right \rangle [P_- -(\gamma-1)\rho_{\rm NS}\varepsilon_{\rm NS}] 
+\rho_{\rm NS}\left \langle R^r_r\right \rangle _-R_{\rm NS}^2\int^{x_s-\epsilon}_{x_{g}}{f(x)x }dx-GM_{\rm NS}\frac{\tau}{\kappa} \, .
\label{eq:midpsi}
\end{eqnarray}
\citet{MURPHY2017} and \citet{Gogilashvili2022} showed that $\Psi \geq 0$ corresponds to the critical condition for explodability. Therefore taking $\Psi=0$ in eq.~(\ref{eq:midpsi}) leads to the following critical condition in the dimensionless form (see details in Appendix \ref{Appendix})
\begin{equation}\label{eq:critcond_eps}
    \tilde{\varepsilon}_{\rm NS} -a \tilde{\kappa} + b\left \langle \tilde{R}^r_r\right \rangle _- = c \, ,
\end{equation}
where $\tilde{\varepsilon}_{\rm NS}$ is the dimensionless specific internal energy measured at $R_{\rm NS}$ and $\tilde{R}^r_r=R^r_r*R_{\rm NS}/(GM_{\rm NS})$ is the dimensionless Reynolds stress.  Coefficient $a$ gives the relative scale of the difference in the pre- and post-shock pressures, coefficient $b$ gives the relative scale of the Reynolds Stress term, and coefficient $c$ gives the relative scale of gravity in this critical condition.  Appendix \ref{Appendix} shows the analytic derivation for each coefficient; we estimate the scales to be $a \approx 0.04$, $b\approx 0.38$, and $c \approx 0.4$. 

To express the critical condition in terms of neutrino power, we next integrate the energy equation, eq.~(\ref{eq:Re}). There are two additional terms due to convection, one is related to the buoyant force, $W_b$, and other to the perturbed internal energy flux, $F_I$ \citep{MABANTA2019}. \citet{MURPHY2011} (2D simulations) and \citet{MURPHY2013} (3D simulations) noticed that buoyant driving roughly balances  turbulent dissipation 
\begin{equation}
    \int_{R_g}^{R_s}W_b dV\approx E_k \, ,
\end{equation}
where $E_k=\int_{R_g}^{R_s}\rho\epsilon_k dV$ is the total power of turbulent dissipation.

 Integrating energy conservation, eq.~(\ref{eq:Re}), leads to an expression for the 
 the specific internal energy in terms of the  neutrino luminosity, $L_\nu$:
 \begin{equation}\label{eq:eps_lum}
    \tilde{\varepsilon}_{\rm NS}=\frac{1}{\gamma}\bigg[\tilde{B}(x_s)+\tilde{L}_\nu \tau_g+\tilde{E}_k+\frac{1}{x_g}\bigg] \, .
\end{equation}
This expressions uses several additional constraints: 1) it uses the analytic boundary condition, eq.~(\ref{eq:jump3}), 2) $\tilde{B}(x_s)$ is the dimensionless Bernoulli integral at the shock, 3) and in the cooling region, neutrino cooling is balanced by neutrino heating and gravitational heating \citep{GOGILASHVILI2021}. $\tilde{E}_k=E_k R_{\rm NS}/(G \dot{M}M_{\rm NS})$ is the dimensionless turbulent energy dissipation rate.
Finally, plugging eq.~(\ref{eq:eps_lum}) into eq.~(\ref{eq:critcond_eps}) gives the FEC in terms of neutrino luminosity:
\begin{equation}\label{eq:FEC+}
    \tilde{L}_\nu \tau_g+\tilde{E}_k-a'\tilde{\kappa}+b'\left \langle \tilde{R}^r_r\right \rangle =c' \, ,
\end{equation}
where $a'=a\gamma$, $b'=b\gamma$, and $c'=c\gamma-\tilde{B}(x_s)-\frac{1}{x_g}$.

Eq.~(\ref{eq:FEC+}) represents an analytic force explosion condition (FEC+) for non-spherical explosions. Compared to the 1D FEC of \citet{Gogilashvili2022}, FEC+ includes the effects of neutrino-driven convection through turbulent dissipation and turbulent ram pressure. Since these convective terms are positive definite, the convection model reduces the neutrino power required for explosion.

To illustrate the quantitative effect of the convection model, we estimate the scale of each term in eq.~(\ref{eq:FEC+}) as follows.  To provide a basis for comparison, the neutrino power near the explosion condition is $\tilde{L}_\nu \tau_g \sim 0.4$ \citep{Gogilashvili2022}, the difference-in-pressures term is $a'\tilde{\kappa}\sim 0.05$. Multi-dimensional simulations indicate that the dimensionless turbulent dissipation is, $\tilde{E}_k\sim 0.12$ and the dimensionless Reynolds stress is $\left \langle \tilde{R}^r_r\right \rangle \sim 0.02$ \citep{MURPHY2013}. Based on these estimates, turbulent dissipation is $\sim 30\%$ of the net neutrino power while turbulent ram pressure term is only $\sim 5\%$ (see appendix~\ref{scales}).  In summary, the convection model reduces the critical condition mostly through turbulent dissipation, and these analytic calculations quantitatively match the numerical results of \citep{MABANTA2018}.

\section{Applications of the FEC+ in different convection models}\label{ssec:Applications}
\subsection{The Integral Convection Model}\label{ssec:MFC}
Eq.~(\ref{eq:FEC+}) represents the FEC+ when considering convection models in general.  This section considers the integral, mean field model of \citet{MURPHY2013}.  While multi-dimensional effects such as neutrino-driven convection and turbulent dissipation are important in the shock revival, simulating two- and three-dimensional CCSN simulations are computationally very expensive, making it difficult to perform systematic studies. Therefore, as it is often the case, the multi-dimensional effects are in some way incorporated into 1D simulations \citep{MURPHY2011,MURPHY2013,Perego2015,MABANTA2019,Couch2020}. One can apply these specific convection models to further simplify the FEC+. 

The integral convection model of \citet{MURPHY2013} starts with fundamental equations of hydrodynamics and Reynolds decomposes physical variables into the background and turbulent parts. Based on multi-dimensional simulations, they show that many of the turbulent correlations are negligible and the ones that are important are the Reynolds stress, $\bm{R}$, the work done by buoyant force, $W_b$, and the turbulent energy flux, $F_I$. To close the system of equations, one needs to know the relation between these quantities. There are several relationships that close the equations.  For one, multi-dimensional simulations \citep{MURPHY2013} indicate that buoyant driving roughly balances turbulent dissipation, $W_b\approx E_k$.  Another relationship is that the  turbulent energy dissipation and the turbulent flux is proportional to the neutrino power absorbed in the gain region: $L_e^{\rm max}\approx \alpha L_\nu \tau$ and $E_k \approx \zeta L_\nu \tau$ \citep{MURPHY2013}. \citet{MABANTA2019} incorporated this integral convection model into 1D simulations and showed that these 1D+ simulations accurately reproduce the explosion conditions of two- and three-dimensional simulations. To relate the Reynolds stress and turbulent dissipation, \citet{MABANTA2019} assume that 
\begin{equation}    
    \epsilon_k \approx \frac{u'^3}{\mathcal{L}}=\frac{R^{r{3/2}}_{r}}{\mathcal{L}} \, ,
    \label{epsk1}
\end{equation}
where $\mathcal{L}$ is the largest turbulent eddy size and $E_k=\int_{r_g}^{r_s}\epsilon_k \rho dV$. On the other hand $\epsilon_k$ can be approximated as 
\begin{equation}
    \epsilon_k=\frac{E_k}{M_g}\, ,
    \label{epsk2}
\end{equation}
where $M_g$  is the mass in the gain region.
Eqs.~(\ref{epsk1}) \&~(\ref{epsk2}) provide a relationship between the Reynolds stress and turbulent dissipation:
\begin{equation}
    R^r_r\approx \bigg(\frac{E_k \mathcal{L}}{M_g}\bigg)^{2/3}\approx \bigg(\frac{\zeta L_\nu \tau_g \mathcal{L}}{M_g}\bigg)^{2/3}
    \label{eq:Rrr}
\end{equation}
 
If one applies these assumptions, the FEC+ (eq.~\ref{eq:FEC+}) becomes:
\begin{equation}
    (1+\zeta)\tilde{L}_\nu \tau_g+b'[\zeta\tilde{\mathcal{L}}]^{2/3}(\tilde{L}_\nu \tau_g)^{2/3}-a'\tilde{\kappa}=c' \, ,
\end{equation}
where the ratio of integrated turbulent dissipation to neutrino power is $\zeta=E_k/(L_\nu \tau_g)\sim 0.3$ and the dimensionless dissipation length-scale is $\tilde{\mathcal{L}}=\mathcal{L}\frac{\dot{M}}{M_g}\sqrt{\frac{R_{\rm NS}}{G M_{\rm NS}}} \approx 7 \times 10^{-5}$.

\subsection{Supernova Turbulence In Reduced-dimensionality (STIR)}\label{ssec:stir}
Another convection model is the STIR model of \citet{Couch2020}.   The STIR method is similar in intent to the integral method in \citet{MABANTA2019}, but there are some important differences.  Both STIR and the integral method in \citet{MABANTA2019} relate the Reynolds stress and turbulent dissipation via eq.~(\ref{epsk1}).  One major difference is that STIR uses a mixing length prescription to close the convection model.  In particular, STIR assumes that the higher order Reynolds correlations are proportional to background gradients.  For example, STIR assumes that the buoyant force is proportional to the Brunt-V\"{a}is\"{a} frequency:
\begin{equation}
    W_b\approx \rho u' \omega_{\rm BV}^2 \Lambda_{\rm mix} \, ,
\end{equation}
where $\omega_{\rm BV}$ is the Brunt-V\"{a}is\"{a}l\"{a} frequency and $\Lambda_{\rm mix}=\alpha_\lambda p/\rho g$ is the mixing length with mixing length parameter $\alpha_\lambda$. Assuming the Ledoux criterion, the Brunt-V\"{a}is\"{a}l\"{a} frequency is
\begin{equation}
    \omega_{\rm BV}^2=-g_{\rm eff}\bigg(\frac{1}{\rho}\frac{\partial \rho}{\partial r}-\frac{1}{\rho c_s^2}\frac{\partial P}{\partial r}\bigg) \, ,
\end{equation}
where $g_{\rm eff}=-\frac{\partial \Phi}{\partial r}+u_r\frac{\partial u_r}{\partial r}$ is effective gravitational acceleration and $c_s$ is the adiabatic sound speed.  Another difference is that STIR includes turbulent production from the divergence of the background flow: $\boldsymbol{\nabla} \cdot (\boldsymbol{u}_0 \left \langle \rho (TrR)\right \rangle )$ which is $\sim u'^2$. This term is ignored both in the integral convection model of \citet{MABANTA2019} as well as in the derivation of the FEC+ in this manuscript (\ref{sec:FEC+}). Another major difference is that \citet{MABANTA2019} assumes steady state and presented an integrated convection model.  In contrast, \citet{Couch2020} evolve the turbulent energy equation.

These differences lead to a different formulation for FEC+. In the following,  we include $\boldsymbol{\nabla} \cdot (\boldsymbol{u}_0 \left \langle \rho (TrR)\right \rangle )$ term and solve for the FEC+. This  additional term in the energy conservation, eq.~(\ref{eq:energy}) leads to an additional term in the FEC+ of $\left \langle R^r_r\right \rangle $. Finally, since STIR solves for the Reynolds stress, a form that is more appropriate for STIR simulations is:
\begin{equation}
    \tilde{L}_{\nu} \tau_g+\frac{\tilde{R}^{r 3/2}_r}{\tilde{\mathcal{L}}}-a'\tilde{\kappa}+(b'+1)\left \langle \tilde{R}^r_r\right \rangle =c'
    \label{eq:FEC+STIR}
\end{equation}

\section{Discussion and Conclusions}\label{sec:discussion}

Understanding the explosion mechanism of CCSNe is important in understanding which massive stars collapse to black holes and which explode.  The 1D FEC of \citet{Gogilashvili2022} accurately predicts the explodability of 1D radiation-hydrodynamic simulations \citep{Gogilashvili2023}. However, CCSN explosions are not spherical, and in fact, multi-dimensional effects such as neutrino-driven convection are important in aiding explosion.  To be be able to predict the outcome of actual CCSN explosions in Nature, we generalize the FEC to include neutrino-driven convection. Eq.~(\ref{eq:FEC+}) presents this more general form, which we call FEC+. Including convection and turbulent dissipation generates two additional terms compared to the 1D FEC \footnote{see eq. (38) in \citet{Gogilashvili2022} and eq.~(\ref{eq:FEC+}) for comparison}: the first is the dimensionless turbulent dissipation, $\tilde{E_k}$; the second is due to turbulent ram pressure, $b'\left \langle \tilde{R}^r_r\right \rangle $. At first glance, there appear to be four dimensionless parameters in FEC+, two additional from the convection model.  However, Kolmogorov's theory for turbulent dissipation relates the Reynolds stress to turbulent dissipation.  This in turn reduces the number of dimensionless parameters. Therefore, the FEC+ effectively depends upon three dimensionless parameters: 1) the dimensionless net neutrino heating deposited in the gain region, 2) the dimensionless neutrino opacity, and 3) the dimensionless Reynolds stress (or turbulent velocity). 

These additional terms due to convection and turbulent dissipation reduce the net neutrino heating required for explosion, and this reduction is consistent with published studies.  The major contributor of this reduction is the turbulent dissipation rate which is $\sim 30\%$ of net neutrino heating deposited in the gain region. The turbulent ram pressure has a smaller effect on the reduction as it is only $\sim 5\%$ of total net neutrino heating.  Taken together, we estimate that the convective model reduces the neutrino heating required for explosion by 26\%.  This analytic estimate is consistent with numerical results.  For one, \citet{MABANTA2018} also found that turbulent dissipation dominates the reduction of the critical condition for explosion.  Moreover, numerical simulations also show that convection reduces the critical condition by $\sim$30\% \citep{MURPHY2011,MURPHY2013,MABANTA2018}.
While these comparisons with previous studies are encouraging, it is important to check the predictions of FEC+ with multi-dimensional CCSN simulations. 

If FEC+ proves to accurately predict the explosion conditions of multi-dimensional simulations, then this theory could significantly improve our understanding of how massive stars explode.  For example, the FEC+ could quantify how close a multi-dimensional simulation is to explosion; it could also help to quantify the relative impact of the numerous physical processes involved.  Additionally, the FEC+ suggests a clear path to predicting which stars explode from the progenitor structure alone, without performing expensive simulations.  The dimensionless parameters depend upon dimension-full parameters such as $L_{\nu}$, $M_{\rm NS}$, $R_{\rm NS}$, and $\dot{M}$.  If one can predict the evolution of these quantities from the progenitor structure alone, then one could determine whether a progenitor structure will explode or not.  Reducing the question of whether real stars explode in Nature to just three dimensionless numbers illustrates the elucidating potential of the FEC+.

\section*{Acknowledgements}
We thank Luca Boccioli for insightful discussions concerning the FEC+ and the STIR convection model. 
This work was supported through the Laboratory Directed Research and Development program, the Center for Space and Earth Sciences, and the center for Nonlinear Studies under project numbers 20240477CR-SES and 20220564ECR at Los Alamos National Laboratory (LANL). 
LANL is operated by Triad National Security, LLC, for the National Nuclear Security Administration of U.S. Department of Energy (Contract No. 89233218CNA000001). 

\appendix
\section{Derivation of the FEC+}
\label{Appendix}
The derivation starts with the fundamental equations of hydrodynamics (eqs.~(\ref{eq:cont})~-~(\ref{eq:energy})). We consider the steady state assumption and Reynolds decompose the physical quantities, $q=q_0+q'$, where $q$ is a generic state variable. The Reynolds decomposed equations are:
\begin{equation}
    (\rho_0 u_0^i)_{;i}+\left \langle \rho' u'^i\right \rangle _{;i}=0
    \label{eq:Rcont}
\end{equation}
\begin{equation}
    (\rho_0 u_{0i} u_0^j+\delta_i^j P_o)_{;j}+\left \langle \rho_0 u'_i u'^j\right \rangle _{;j}+\left \langle \rho' u_{0i} u'^j\right \rangle _{;j}=-\rho_0 \Phi_{,i}
\end{equation}
\begin{eqnarray}\label{eq:Renergy}
    &(\rho_0 u_0^j(\varepsilon_0+\frac{u_0^2}{2}))_{;j}+\left \langle \rho_0 u'^j \varepsilon'\right \rangle _{;j}+\left \langle \rho_0 u_0^j\frac{u'^2}{2}\right \rangle _{;j} +\left \langle \rho_0u'^ju_{0k}u'^k\right \rangle _{;j}& \\ \nonumber
    &\left \langle \rho' u_0^j \varepsilon'\right \rangle _{;j}+\left \langle \rho' u_0^j u_{0k} u'^k\right \rangle _{;j}+\left \langle \rho'u'^j\varepsilon_0\right \rangle _{;j}+\left \langle \rho'u'^j\frac{u0^2}{2}\right \rangle _{;j}& \\ \nonumber
    &+ (u_0^j P_0)_{;j}+\left \langle (u'^j P')\right \rangle _{;j}=-(\rho_0 u_0^j \Phi_{,j})-\left \langle \rho'u'^j\right \rangle \Phi_{,j}+\rho_0 q& \, ,
\end{eqnarray}
where $\left \langle \cdot \right \rangle$ is an average defined over the solid angle and time.
Multi-dimensional simulations show that many of the terms in eqs.~(\ref{eq:Rcont})~-~(\ref{eq:Renergy}) are negligible \citep{MURPHY2011,MURPHY2013}.  These terms are small due to the Boussinesq approximation, in which density perturbations are small for all terms except the buoyant driving term.  In addition, the pressure perturbations, $P'$, are small due to being higher order perturbations. The resulting equations are the eqs.~(\ref{eq:Rc})~-~(\ref{eq:Re}).

Similar to \citet{Gogilashvili2022}, we then integrate eqs.~(\ref{eq:Rc})~-~(\ref{eq:Re}) from the NS to the shock. \citet{Gogilashvili2022} derive the $\Psi$ parameter from the (im)balance of forces in the momentum conservation equation, eq.~(\ref{eq:Rm}). Therefore, let's first consider the momentum equation.
\begin{eqnarray}
&\int^{R_s+\epsilon}_{R_{\rm NS}}{\frac{1}{r^2}\frac{\partial}{\partial r}(r^2 \rho_0 u_{0r}^2) dV} + \int^{R_s+\epsilon}_{R_{\rm NS}}{\frac{\partial P_0}{\partial r}dV}+\int^{R_s+\epsilon}_{R_{\rm NS}}[\frac{1}{r^2}\frac{\partial}{\partial r}\left \langle r^2\rho_0 R^r_r\right \rangle &  \nonumber \\
&-\frac{1}{r}\left \langle \rho_0R^\theta_\theta+\rho_0R^\phi_\phi\right \rangle ]dV=\int^{R_s+\epsilon}_{R_{\rm NS}}{\rho_0 \frac{\partial \Phi}{\partial r}dV}&
\end{eqnarray}
Integrating to $R_s + \epsilon$ includes the shock conditions in the equations. To eliminate the jump conditions, we then split these integrals into the integral up to shock and at the shock: $\int_{R_{\rm NS}}^{R_s+\epsilon}=\int_{R_{\rm NS}}^{R_s-\epsilon}+\int_{R_{s}-\epsilon}^{R_s+\epsilon}$. This allow us to eliminate the boundary terms. 

\begin{eqnarray}
\centering
    &&\int^{R_s-\epsilon}_{R_{\rm NS}}{\frac{1}{r^2}\frac{\partial}{\partial r}(r^2 \rho_0 u_{0r}^2) dV} + \int^{R_s-\epsilon}_{R_{\rm NS}}{\frac{\partial P_0}{\partial r}dV}+\int^{R_s-\epsilon}_{R_{\rm NS}}[\frac{1}{r^2}\frac{\partial}{\partial r}\left \langle r^2\rho_0 R^r_r\right \rangle   \nonumber \\
&-&\frac{1}{r}\left \langle \rho_0R^\theta_\theta+\rho_0R^\phi_\phi\right \rangle ]dV + \cancelto{Boundary~Condition}{\int^{R_s+\epsilon}_{R_s-\epsilon}{[\frac{\partial}{\partial r}(r_0u_{0r}^2+P_0+\left \langle \rho_0 R^r_r\right \rangle )]dV}}\\ \nonumber
&+&\cancelto{Small (\sim \epsilon)}{\int^{R_s+\epsilon}_{R_s-\epsilon}{[\frac{2}{r}\rho_0 u_{0r}^2 +\frac{2}{r}\left \langle \rho_0R^r_r\right \rangle }-\frac{1}{r}\left \langle \rho_0R^\theta_\theta+\rho_0 R^\phi_\phi\right \rangle ]dV}=-\int^{R_s-\epsilon}_{R_{\rm NS}}{\rho_0\frac{\partial\Phi}{\partial r}dV} \\ \nonumber
&-&\cancelto{Small (\sim \epsilon)}{\int^{R_s+\epsilon}_{R_s-\epsilon}{\rho_0 \frac{\partial \Phi}{\partial r}dV}} 
\end{eqnarray}
Therefore, the resulting $\Psi$ parameter is 
\begin{eqnarray}\label{eq:psiappend}
    \Psi&=&-\int^{R_s-\epsilon}_{R_{\rm NS}}{\frac{1}{r^2}\frac{\partial}{\partial r}(r^2 \rho_0 u_{0r}^2) dV} - \int^{R_s-\epsilon}_{R_{\rm NS}}{\frac{\partial P_0}{\partial r}dV}-\int^{R_s-\epsilon}_{R_{\rm NS}}[\frac{1}{r^2}\frac{\partial}{\partial r}\left \langle r^2\rho_0 R^r_r\right \rangle  \nonumber \\
&-&\frac{1}{r}\left \langle \rho_0R^\theta_\theta+\rho_0R^\phi_\phi\right \rangle ]dV -\int^{R_s-\epsilon}_{R_{\rm NS}}{\rho_0\frac{\partial\Phi}{\partial r}dV}
\end{eqnarray}
Compared to $\Psi$ for the spherically symmetric case \citep{Gogilashvili2022}, there are only two additional terms due to convection, the third and fourth terms in eq.~(\ref{eq:psiappend}).  CCSN simulations show that in neutrino-driven convection, there is an equipartition between the radial direction and both tangential directions of Reynolds stress, $R_{rr}\sim R_{\theta\theta}+R_{\phi \phi}$. In addition, the off-diagonal terms of the Reynolds stress are small \citep{MURPHY2013}.  This relation between the radial and tangential components of Reynolds stress allows us to further simplify the additional terms in the momentum equation, in particular:  $-\frac{1}{r}\left \langle \rho_0R^\theta_\theta+\rho_0R^\phi_\phi\right \rangle \approx -\frac{1}{r}\left \langle \rho_0 R^r_r\right \rangle $ \citep{MURPHY2013}. These same simulations also show that $\left \langle R^r_r\right \rangle $ is roughly constant in the gain region and quickly tapers to zero at the base of the gain region such that $\left \langle R^r_r\right \rangle $ vanishes at the NS "surface".  At the upper end of the convective region, $\left \langle R^r_r\right \rangle $ remains nonzero all the way up to the shock.  The density profile scale by the density at the NS surface is $\rho=\rho_{\rm NS}f(x)$. These assumptions allow us to rewrite the integrals in terms of relevant scales such as $\rho_{\rm NS}$ and an average Reynolds stress in the gain region.  This in turn facilitates an algebraic and analytic condition.  The nuances in the profile are absorbed into an integral of the dimensionless function $f(x)$. Combing these assumptions, the third and fourth terms are:

\begin{equation}
4\pi\int^{R_s-\epsilon}_{R_{\rm NS}}{\frac{\partial}{\partial r}(r^2 \rho_0\left \langle R^r_r\right \rangle ) dr}=4\pi R_s^2\rho_0\left \langle R^r_r\right \rangle _-
\end{equation}
\begin{equation}
   -\int^{R_s-\epsilon}_{R_{\rm NS}}\frac{1}{r} \rho_0\left \langle R^r_r\right \rangle dV=-4\pi\rho_{\rm NS}R_{\rm NS}^2\left \langle R^r_r\right \rangle _-\int^{x_s-\epsilon}_{x_{g}}{f(x)x }dx
\end{equation}

Dropping subscripts ``0'' and ``r'' and using boundary conditions, the third and fourth terms become:
\begin{eqnarray}\label{eq:appendpsif}
    &&\frac{|\dot{M}|}{4\pi \beta}\sqrt{\frac{2GM_{\rm NS}}{R_s}}+\left \langle r^2\right \rangle \bigg( 1-\frac{1}{\beta}\bigg)\frac{|\dot{M}|}{4\pi R_s^2}\sqrt{\frac{2GM_{\rm NS}}{R_s}}\\ \nonumber
    &&+(R_s^2-\left \langle r^2\right \rangle )\rho_{\rm NS}f(x_s)\left \langle R^r_r\right \rangle _- -\left \langle r^2\right \rangle (\gamma-1)\rho_{\rm NS}\varepsilon_{\rm NS}+GM_{\rm NS}\frac{\tau}{\kappa}\\ \nonumber
    &&-\rho_{\rm NS}R_{\rm NS}^2\left \langle R_r^r\right \rangle _-\int^{x_s-\epsilon}_{x_{g}}{f(x)x }dx=0 \, ,
\end{eqnarray}
We normalize these terms in the same way as \citet{Gogilashvili2022} (eqs. (7)-(12) in \citet{Gogilashvili2022}). We also make the same assumptions as \citet{Gogilashvili2022}; for example, we assume that $\dot{M}$ is constant in the gain region $|\dot{M}|=4\pi r_s^2 \rho_+u_+$ and that the preshock material is cold and falls onto the stalled shock at free-fall velocity.

Finally, we assume that $\Psi \ge 0$ describes the explosive conditions and
from eq.~(\ref{eq:appendpsif}), derive the critical condition for the dimensionless specific internal energy:
\begin{eqnarray}
    \tilde{\varepsilon}_{\rm NS}&-& \frac{\bigg[ \frac{\sqrt{2}}{4\pi \beta}x_s^{-1/2}+\frac{\left \langle x^2\right \rangle }{4\pi}(1-\frac{1}{\beta})\sqrt{2}x_s^{-3/2}\bigg] F(x_s)}{\left \langle x^2\right \rangle (\gamma-1)\tau}\tilde{\kappa} \\ \nonumber
    &+& \frac{[-(x_s^2-\left \langle x^2\right \rangle )f(x_s)+\int^{x_s}_{x_g}f(x)xdx]}{\left \langle x^2\right \rangle (\gamma-1)}\left \langle \tilde{R}^r_r\right \rangle _- \\ \nonumber
    &=& \frac{F(x_s)}{\left \langle x^2\right \rangle (\gamma-1)}
\end{eqnarray}
or 
\begin{equation}
    \tilde{\varepsilon}_{\rm NS} -a \tilde{\kappa} + b\left \langle \tilde{R}^r_r\right \rangle _- = c \, ,
\end{equation}
where 
\begin{equation}
    a= \frac{\bigg[ \frac{\sqrt{2}}{4\pi \beta}.x_s^{-1/2}+\frac{\left \langle x^2\right \rangle }{4\pi}(1-\frac{1}{\beta})\sqrt{2}x_s^{-3/2}\bigg] F(x_s)}{\left \langle x^2\right \rangle (\gamma-1)\tau}\, ,
\end{equation}
\begin{equation}
    b=\frac{[-(x_s^2-\left \langle x^2\right \rangle )f(x_s)+\int^{x_s}_{x_g}f(x)xdx]}{\left \langle x^2\right \rangle (\gamma-1)}\, ,
\end{equation}
\begin{equation}
    c=\frac{F(x_s)}{\left \langle x^2\right \rangle (\gamma-1)}\, ,
\end{equation}
where $F(x_s)=\int_{1}^{x_s} f(x)dx$.

To derive the critical condition for the neutrino luminosity, we consider the energy conservation equation, eq.~(\ref{eq:Re}). Using boundary condition, eq.~(\ref{eq:jump3}), we have
\begin{equation}
    [\dot{M}B(x_s)-4\pi R_s^2 F_I(x_s)]-\dot{M}\bigg[\gamma\varepsilon_{\rm NS}-\frac{GM_{\rm NS}}{R_{\rm NS}}\bigg]-\int^{R_s}_{R_{\rm NS}}\rho q dV= -\int^{R_s}_{R_{\rm NS}} \left \langle \rho'u'^j\right \rangle \Phi_{,j}dV-\int^{R_s}_{R_{\rm NS}} \left \langle \rho \varepsilon u'^j\right \rangle _{;j}dV \, ,\nonumber
\end{equation}
\citet{MURPHY2011} found that the Buoyant driving is mostly balanced by the turbulent dissipation

\begin{equation}
    -\int^{R_s}_{R_{\rm NS}} \left \langle \rho'u'^j\right \rangle \Phi_{,j}dV = -\int^{R_s}_{R_g} \left \langle \rho'u'^j\right \rangle \Phi_{,j}dV  \approx E_k
\end{equation}
and 
\begin{equation}
    -\int^{R_s}_{R_{\rm NS}} \left \langle \rho \varepsilon u'^j\right \rangle _{;j}dV = -\int^{R_s}_{R_g} \left \langle \rho \varepsilon u'^j\right \rangle _{;j}dV \approx -4\pi R_s^2 F_I(x_s)= -L_e^{\rm max}
\end{equation}
From there we get eq.~(\ref{eq:eps_lum}) which leads us to the critical condition for neutrino luminosity, eq.~(\ref{eq:FEC+}).

\section{Calculating the scales of the FEC+} \label{scales}

Eq.~(\ref{eq:FEC+}) is generalized explosion conditon with convection and turbulent dissipation. These multi-dimensional effects are demonstrated in second and fourth terms in eq.~(\ref{eq:FEC+}). To investigate the quantitative effect of convection and turbulent dissipation on the explosion condition, we need to estimate the scales of the additional terms with respect to net neutrino power. First, note that the scale of the dimensionless neutrino power is $\tilde{L}_\nu \tau_g \sim 0.4$ \citep{Gogilashvili2022}. From multi-dimensional simulations, the scales of the Reynolds stress and turbulent dissipation are, $R^r_r \lesssim 7\times 10^{17} \rm erg/g$ and $E_k\sim 0.5\times 10^{51} \rm egr/s$ respectively \citep{MURPHY2013}. The dimensionless Reynolds stress and turbulent dissipation are:
\begin{equation}
    \tilde{R}^r_r=R^r_r \frac{R_{\rm NS}}{GM_{\rm NS}}
\end{equation}
and 
\begin{equation}
    \tilde{E}_k=E_k\frac{R_{\rm NS}}{G |\dot{M}| M_{\rm NS}}
\end{equation}

For a typical mass of the NS, $M_{\rm NS}\sim 1.4 M_\odot$, the radius of the NS, $R_{\rm NS}\sim50 \rm km$, and mass accretion rate, $|\dot{M}|\sim 0.05M_\odot$, the dimensionless Reynolds stress and dimensionless turbulent dissipation are, $\tilde{R}^r_r\lesssim 0.02$ and $\tilde{E}_k\sim 0.12$. Therefore, $\tilde{E}_k/\tilde{L}_\nu \tau_g\sim 30\%$ while $\tilde{R}^r_r/\tilde{L}_\nu \tau_g \lesssim 5\%$. Hence, the the reduction of the critical neutrino power is mostly due to turbulent dissipation. The contribution of the turbulent ram pressure is small.

\bibliography{References}{}
\bibliographystyle{aasjournal}

\end{document}